\documentclass[aps,preprint]{revtex4}%
\usepackage{amsfonts}
\usepackage{amsmath}
\usepackage{amssymb}
\usepackage{graphicx}%
\providecommand{\U}[1]{\protect\rule{.1in}{.1in}}
\begin{document}
\preprint{ }
\title[spontaneous emission]{Angular momentum conservation in two-atom superradiance}
\thanks{Contact: P. R. Berman, pberman@umich.edu}
\author{P. R. Berman}
\affiliation{Physics Department, University of Michigan, Ann Arbor, MI \ 48109-1040}
\author{Zeyuan Wang}
\affiliation{Physics Department, University of Michigan, Ann Arbor, MI \ 48109-1040}
\author{A. Kuzmich}
\affiliation{Physics Department, University of Michigan, Ann Arbor, MI \ 48109-1040}
\keywords{spontaneous emission, angular momentum, field energy, rotating-wave
approximation,Weisskopf-Wigner approximation}
\pacs{PACS number}

\begin{abstract}
Two-atom superradiance is studied for atoms having a $J=0$ ground state and a
$J=1$ excited state. The atoms lie on the $x-$axis and are excited by a
$\sigma_{+}$ polarized laser pulse propagating in the $z-$direction. We use
two, complementary approaches to obtain the final angular momentum in the
field, a source-field approach and a Schr\"{o}dinger equation approach. Both
approaches give identical answers, but it is necessary to include terms in
both calculations that, at first glance, might not be expected to contribute.
We show that, owing to an exchange between a state in which (atom 1 is in its
$m=1$ excited state and atom 2 in its ground state) to one in which (atom 2 is
in its $m=-1$ excited state and atom 1 in its ground state), the final angular
momentum in the field is not necessarily equal to the initial angular momentum
in the two-atom system. The missing angular momentum appears as orbital
angular momentum of the two atoms.

\end{abstract}
\volumeyear{year}
\volumenumber{number}
\issuenumber{number}
\eid{identifier}
\date[Date text]{date}
\received[Received text]{date}

\revised[Revised text]{date}

\accepted[Accepted text]{date}

\published[Published text]{date}

\startpage{1}
\maketitle

\section{\bigskip Introduction}

In a recent paper \cite{ber}, a general expression was derived for the
radiation emitted by two atoms, each having a ground state with angular
momentum $J=0$ and an excited state with angular momentum $J=1$. The atoms
were prepared in their fully inverted state by a $\sigma_{+}$ polarized field
incident in the $z$ direction, such that each atom was initially in its $m=1$
excited-state sublevel. This is the simplest example of Dicke superradiance
\cite{dicke} from a fully inverted initial state. The degeneracy of the
excited state played a critical role in determining the specific form of the
radiation pattern, as did the relative orientation of the two atoms.

It is not difficult to understand why the radiation pattern of the two-atom
system depends on the relative orientation of the atoms. The atoms interact
via the vacuum radiation field which couples a state in which atom 1 is
excited and atom 2 in its ground state to a state in which atom 2 is excited
and atom 1 in its ground state (the much weaker van der Waals interaction
between two excited atoms or two ground state atoms is neglected). Suppose the
atoms are in a state in which atom 1 is in its $m=1$ excited-state sublevel
and atom 2 in its ground state. If the atoms lie on the $z-$axis, this state
can be coupled only to the state in which in which atom 2 is in its $m=1$
excited-state sublevel and atom 1 in its ground state. On the other hand, if
the atoms lie on the $x-$axis, this state can also be coupled to one in which
atom 2 is in its $m=-1$ excited-state sublevel and atom 1 in its ground state.
In Ref. \cite{ber}, these exchanges resulted in a time-integrated radiation
pattern that is identical to that for non-interacting atoms if the atoms lie
on the $z-$axis, but not if the atoms lie on the $x-$axis. Although not
calculated explicitly in that paper, it can be shown easily that the
expectation value of the energy in the field after the atoms decay is equal to
the expectation value of the initial energy in the atoms.

An exchange in which excitation is transferred from one magnetic state
sublevel in atom 1 to a different magnetic state sublevel in atom 2 does not
conserve overall angular momentum in the internal states of the 2-atom system.
As a consequence, it is to be expected that such an exchange will have an
effect on the final angular momentum of the field radiated by the two atoms as
they decay. In this paper, we calculate the total angular momentum in the
field following the decay of the atoms. For a completely inverted initial
state in which each atom is initially in its $m=1$ excited-state sublevel, the
expectation value of the initial angular momentum in the two-atom system is
$\left\langle \mathbf{L}_{atoms}(t=0)\right\rangle =2\hbar\mathbf{\hat{z}}$.
Following the decay, $\left\langle \mathbf{L}_{atoms}(t=\infty)\right\rangle
=\mathbf{0}$. If the atoms are on the $z-$axis, the final angular momentum in
the field is equal to $2\hbar\mathbf{\hat{z}}$; that is, $\left\langle
\mathbf{L}_{field}(t=\infty)\right\rangle =2\hbar\mathbf{\hat{z}}$ and angular
momentum is conserved for the atom-field system. On the other hand, if the
atoms are on the $x-$axis, this is no longer the case, $\left\langle
\mathbf{L}_{field}(t=\infty)\right\rangle \neq2\hbar\mathbf{\hat{z}}$, owing
to the the exchange of magnetic state sublevels as the atoms decay. Where did
the angular momentum go? We will show that the missing angular momentum goes
into \textit{orbital} angular momentum of the two-atom system \cite{orb}. To
simplify the calculation we assume that the atoms are infinitely massive,
allowing us to avoid problems with any motion of the atoms as they decay. The
calculations are restricted to the case when the two atoms lie on the
$x-$axis. For atoms on the $z-$axis, there is no torque exerted by one atom on
the other and the initial angular momentum in the atoms is transferred totally
to the radiated field.

If the atoms are separated my much less than a wavelength, it will turn out
that $\left\langle \mathbf{L}_{field}(t=\infty)\right\rangle =\hbar
\mathbf{\hat{z}}$. For such separations, there is a rapid transfer between the
$m=1$ and $m=-1$ sublevels of the excited states of the atoms following the
emission of the first photon. As a consequence of this exchange, as the second
photon is emitted, there is no angular momentum transferred to the radiation
field . To further understand this phenomena we also consider an initial
condition in which atom 1 is in its $m=1$ excited state sublevel and atom 2 in
its ground state. In this case, owing to the rapid exchange between $m=1$ and
$m=-1$ sublevels, the final angular momentum in the radiated field is equal to
zero for atoms separated by much less than a wavelength.

In Sec. II we introduce the atom-field geometry and define the molecular
states. A source-field approach similar to the one used in Ref. \cite{ber} is
then used in Sec III to calculate the final angular momentum in the field. In
Sec. IV, we calculate the change in orbital angular momentum of the two-atom
system and show that it accounts for he difference between the initial angular
momentum in the atoms and the final angular momentum in the field. To provide
an alternative method for calculating the field angular momentum, in Sec. V,
we use the Schr\"{o}dinger picture approach proposed in Ref. \cite{jmo} to
calculate the angular momentum in the field. In order to illustrate the
relevant physics, we consider only two atoms and a $J=1$ to $J=0$ transition.
Modifications of the theory neede to generalize the calculation to transitions
between states having arbitrary angular momentum and to systems of $N$ atoms
are discussed briefly in Sec. VI.

Some changes to the formalisms used in Refs. \cite{ber} and \cite{jmo} must be
incorporated into the calculation of the angular momentum of the fields
radiated by the two-atom system. That is, the conventional source-field theory
used in Ref. \cite{ber} to calculate the fields in the radiation zone must be
modified to include corrections arising from the finite displacement of the
two atoms and a term that can be neglected in the calculation of the angular
momentum in the field radiated by a single atom given in Ref. \cite{jmo} must
now be retained.

There are several papers on the angular momentum in the field radiated by a
single atom \cite{jmo,angpap} and there are hundreds, if not thousands of
papers on Dicke superradiance, but we are unaware of any papers that focus on
the angular momentum of the fields radiated by the two-atom system that we
consider in this paper. As such, this paper provides the first analysis of the
angular momentum conservation in two-atom Dicke superradiance, a problem of
fundamental importance in quantum optics. Although we use some of the
formalism of Ref. \cite{ber}, the calculation differs considerably from that
of Ref. \cite{ber}.

\section{\bigskip General Considerations}

\begin{figure}[ptb]
\centering
\includegraphics[width=\textwidth,height=\textheight,keepaspectratio
]{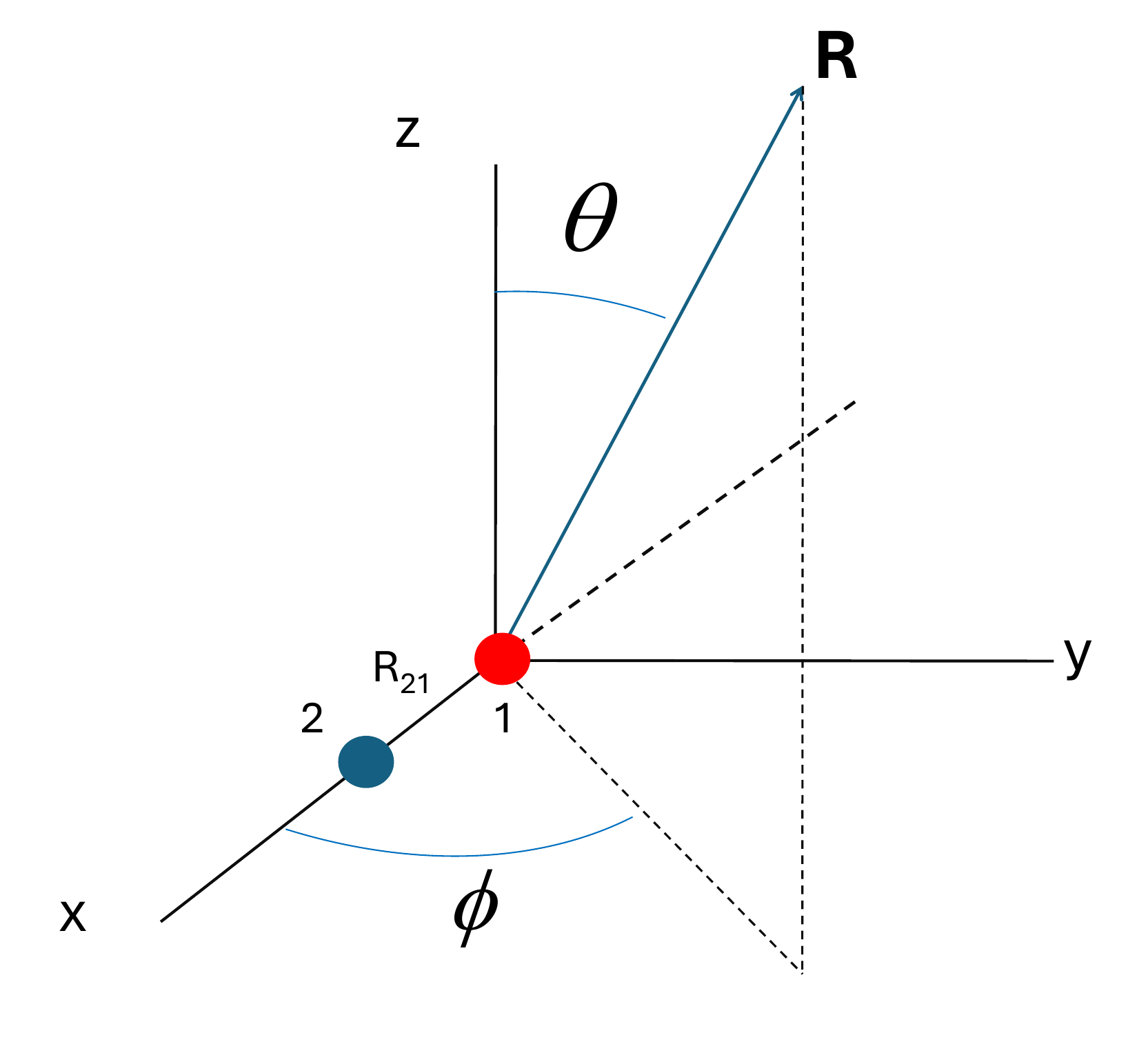} \caption{A schematic representation of the geometry used in the
calculation. The radiated fields are evaluated at a distance $R$ \ from the
origin, with polar angle $\theta$ and azimuthal angle $\phi$. Atom 1 (red dot)
is at the origin and atom 2 is on the $x-$axis (blue dot), a distance $R_{21}$
from the origin.}%
\label{angfig1}%
\end{figure}

Atom 1 is located at the origin and atom 2 at $\mathbf{R}_{2}=R_{21}%
\mathbf{\hat{x}}$ (see Fig. \ref{angfig1}). The fields are calculated at a
distance $R$ from the origin with polar angle $\theta$ and azimuthal angle
$\phi$. The atoms are excited using a $\sigma_{+}$ polarized pulse propagating
in the $z-$direction. Two initial conditions are considered, referred to as DE
(double excitation) and SE (single excitation). For the DE initial condition,
both atoms are in their $m=1$ excited state sublevels at $t=0$ and for the SE
initial condition, atom 1 is in its $m=1$ sublevel and atom 2 in its ground
state at $t=0$. It is assumed that the temporal envelope of the pulse is
sufficiently small to allow for "instantaneous" excitation of the atoms into
the initial state.

%EndExpansion

The atoms are identical, each having a $J=0$ ground state $\left\vert
g\right\rangle $ and a $J=1$ excited state, whose three sublevels have an
associated ket $\left\vert m\right\rangle $, with $m=0,\pm1$. The ground to
excited state transition frequency in each atom is equal to $\omega_{0}%
=k_{0}c$. For \textit{noninteracting }atoms the eigenkets can be written as
$\left\vert \alpha\alpha^{\prime}\right\rangle =\left\vert \alpha\right\rangle
\left\vert \alpha^{\prime}\right\rangle $, where the first symbol refers to a
state of atom 1 and the second to a state of atom 2. For example, $\left\vert
g1\right\rangle $ represents an eigenket in which atom 1 is in its ground
state and atom 2 in its $m=1$ excited state.

\begin{figure}[ptb]
\centering
\includegraphics[width=\textwidth,height=\textheight,keepaspectratio
]{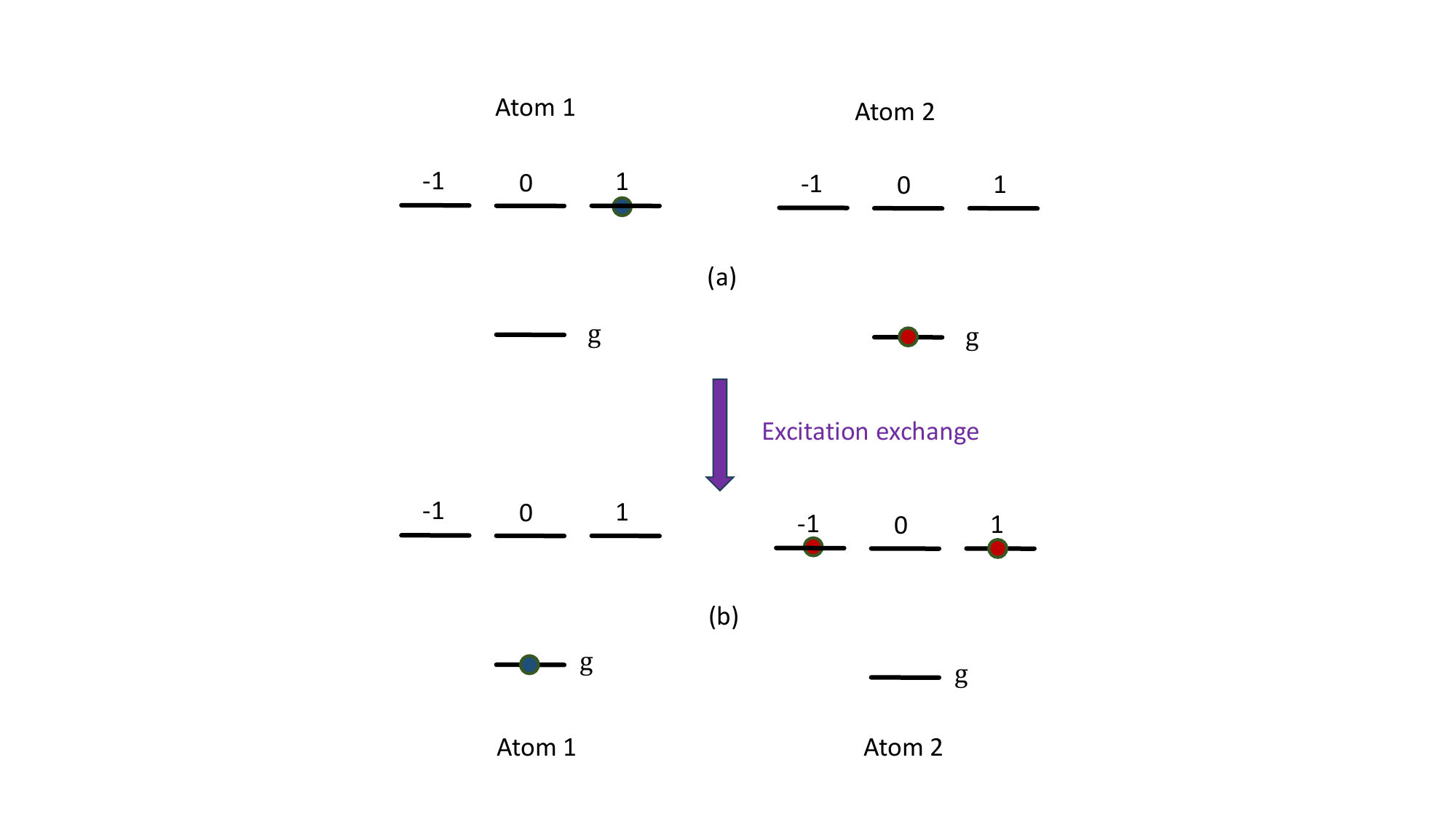}\caption{Schematic representation of the excitation exchange
process. (a) Atom 1 is initially in its $m=1$ excited state sublevel and atom
2 in its ground state. (b) As a result of excitation exchange for atoms on the
$x-$axis, atom 2 can be excited into a linear superposition of its $m=\pm1$
sublevels while atom 1 is returned to its ground state. The transfer to the
$m=-1$ excited state sublevel results in a torque exerted on atom 2 by atom
1.}%
\label{angfig1a}%
\end{figure}

The atoms interact via the vacuum radiation field. With the neglect of van der
Waals interactions between two excited atoms or two ground state atoms, the
only states that are coupled by the vacuum radiation field are $\left\vert
gm\right\rangle $ and $\left\vert m^{\prime}g\right\rangle $, with
$m,m^{\prime}=0,\pm1$. In other words, the vacuum field interaction results in
an \textit{exchange} between a magnetic state sublevel in one atom with the
same or another magnetic state sublevel in the other atom. For our initial
conditions, the only states involving a single excitation that can be
populated are $\left\vert g,\pm1\right\rangle $ and $\left\vert \pm
1,g\right\rangle $. The exchange interaction is depicted schematically in Fig.
\ref{angfig1a}. Owing to this exchange coupling, the "bare-atom" basis is not
particularly convenient for carrying out the calculations.

%EndExpansion

To understand why this is the case, we can examine the atom-field dynamics in
the bare-atom basis. If there were no excitation exchange and there was at
most one excitation in the two-atom system, the excited state amplitudes and
$b_{gm}$ and $b_{mg}$ would undergo simple exponential decay,
\begin{equation}
b_{mg}(t)=e^{-\gamma_{2}t/2}b_{mg}(0)\text{; \ \ \ \ \ }b_{gm}(t)=e^{-\gamma
_{2}t/2}b_{gm}(0),
\end{equation}
where%
\begin{equation}
\gamma_{2}=\frac{\mu_{12}^{2}\omega_{0}^{3}}{9\hbar\pi\epsilon_{0}c^{3}}%
\end{equation}
is the ($m-$independent) excited state decay rate of an isolated atom and
$\mu_{12}$ is a reduced matrix element (assumed real) associated with the
ground to excited state transition. However, when excitation exchange is
included, the state amplitudes no longer decay exponentially. In the Appendix,
we derive the evolution equations for the state amplitudes using both the
rotating wave approximation (RWA) and the Weisskopf-Wigner approximation (WWA)
\cite{ww}. These equations, given in Eqs. (\ref{ev}), exhibit both the
vacuum-field induced spontaneous decay of the excited states and the
vacuum-field induced excitation exchange between the state amplitudes
$b_{\pm1g}(t)$ and $b_{g,\pm1}(t)$.

\begin{figure}[ptb]
\centering
\includegraphics[width=\textwidth,height=\textheight,keepaspectratio
]{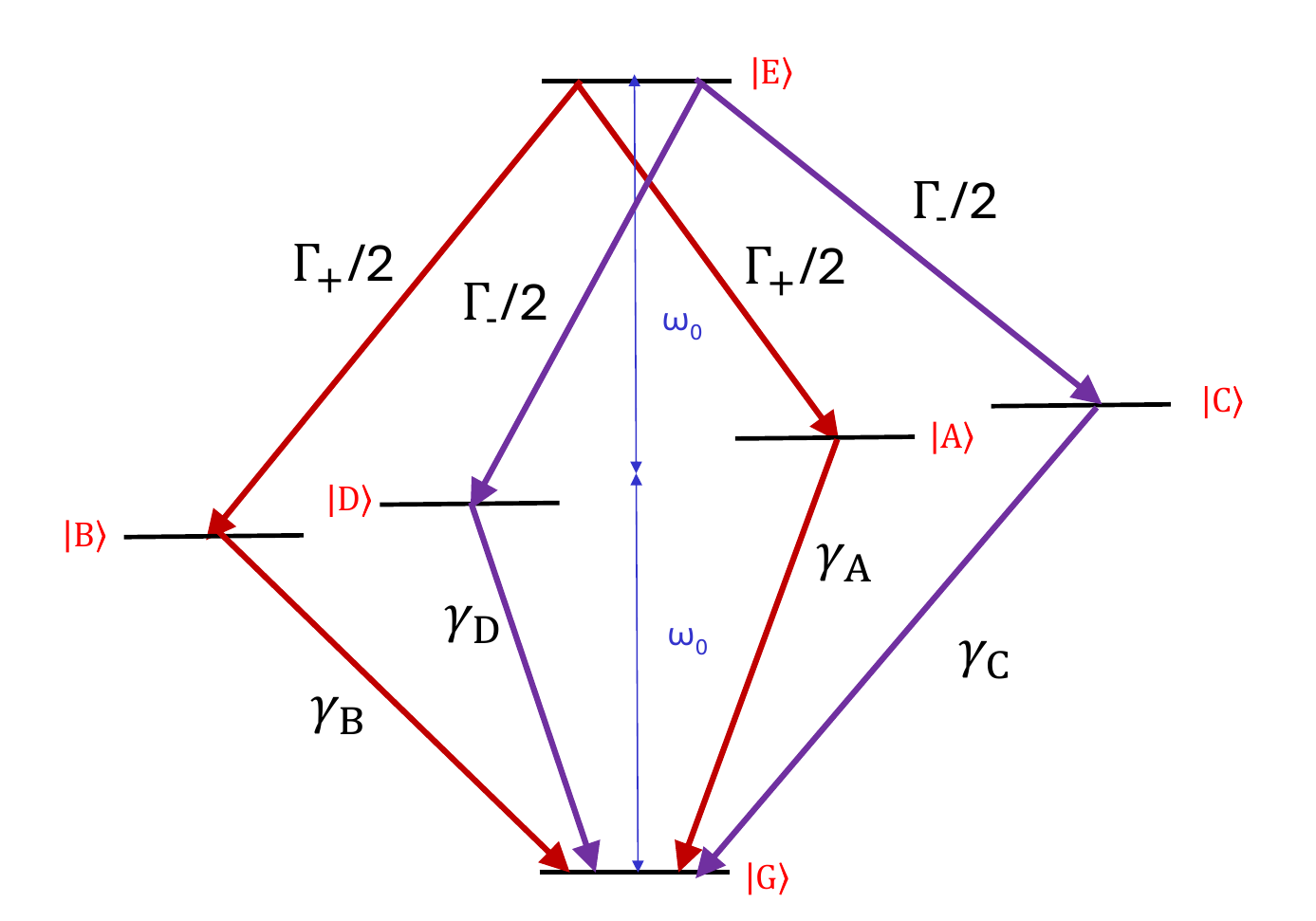}\caption{Molecular basis level scheme. The intermediate state
eigenfrequencies are \ $\omega_{A}=\omega_{0}+\gamma_{2}\left(  q_{a}%
+q_{b}\right)  /2$, $\omega_{B}=\omega_{0}+\gamma_{2}\left(  q_{a}%
-q_{b}\right)  /2$, $\omega_{C}=\omega_{0}-\gamma_{2}\left(  q_{a}%
-q_{b}\right)  /2,$ and $\omega_{D}=\omega_{0}-\gamma_{2}\left(  q_{a}%
+q_{b}\right)  /2$. In addition to populating the intermediate states, decay
from state $\left\vert E\right\rangle $ also creates the coherences $\rho
_{AB},$ $\rho_{BA}$, $\rho_{CD},$ and $\rho_{DC}$.}%
\label{angfig2}%
\end{figure}

To obtain state amplitudes that undergo simple exponential decay, we can
diagonalize Eqs. (\ref{ev}). In doing so, we obtain the "molecular state"
eigenkets of the combined two-atom system given by
\begin{subequations}
\label{molbas}%
\begin{align}
\left\vert G\right\rangle  &  =\left\vert gg\right\rangle ,\\
\left\vert A\right\rangle  &  =\frac{\left\vert 1g\right\rangle +\left\vert
g1\right\rangle +\left\vert -1,g\right\rangle +\left\vert g,-1\right\rangle
}{2},\\
\left\vert B\right\rangle  &  =\frac{\left\vert 1g\right\rangle +\left\vert
g1\right\rangle -\left\vert -1,g\right\rangle -\left\vert g,-1\right\rangle
}{2},\label{79}\\
\left\vert C\right\rangle  &  =\frac{\left\vert 1g\right\rangle -\left\vert
g1\right\rangle -\left(  \left\vert -1,g\right\rangle -\left\vert
g,-1\right\rangle \right)  }{2},\\
\left\vert D\right\rangle  &  =\frac{\left\vert 1g\right\rangle -\left\vert
g1\right\rangle +\left(  \left\vert -1,g\right\rangle -\left\vert
g,-1\right\rangle \right)  }{2},\\
\left\vert E\right\rangle  &  =\left\vert 11\right\rangle .
\end{align}
These molecular-state eigenkets are indicated schematically in Fig.
\ref{angfig2}. The associated eigenfrequencies and decay rates are given by
\cite{ber}
\end{subequations}
\begin{subequations}
\label{freqpar}%
\begin{align}
\omega_{G}  &  =0,\\
\omega_{A}  &  =\omega_{0}+\gamma_{2}\left[  q_{a}\left(  x_{0}\right)
+q_{b}\left(  x_{0}\right)  \right]  /2,\\
\omega_{B}  &  =\omega_{0}+\gamma_{2}\left[  q_{a}\left(  x_{0}\right)
-q_{b}\left(  x_{0}\right)  \right]  /2,\\
\omega_{C}  &  =\omega_{0}-\gamma_{2}\left[  q_{a}\left(  x_{0}\right)
-q_{b}\left(  x_{0}\right)  \right]  /2,\\
\omega_{D}  &  =\omega_{0}-\gamma_{2}\left[  q_{a}\left(  x_{0}\right)
+q_{b}\left(  x_{0}\right)  \right]  /2,\\
\omega_{E}  &  =2\omega_{0},
\end{align}
\end{subequations}
\begin{subequations}
\label{decpar}%
\begin{align}
\Gamma_{\pm}  &  =\gamma_{2}\left[  1\pm p_{a}\left(  x_{0}\right)  \right] \\
\gamma_{A}  &  =\gamma_{2}\left[  1+p_{a}\left(  x_{0}\right)  +p_{b}\left(
x_{0}\right)  \right]  ,\\
\gamma_{B}  &  =\gamma_{2}\left[  1+p_{a}\left(  x_{0}\right)  -p_{b}\left(
x_{0}\right)  \right]  ,\\
\gamma_{C}  &  =\gamma_{2}\left[  1-p_{a}\left(  x_{0}\right)  +p_{b}\left(
x_{0}\right)  \right]  ,\\
\gamma_{D}  &  =\gamma_{2}\left[  1-p_{a}\left(  x_{0}\right)  -p_{b}\left(
x_{0}\right)  \right]  ,
\end{align}
where
\end{subequations}
\begin{subequations}
\label{pq}%
\begin{align}
p_{a}\left(  x_{0}\right)   &  =\frac{3}{4}\left[  \frac{\sin x_{0}}{x_{0}%
}-\frac{\cos x_{0}}{x_{0}^{2}}+\frac{\sin x_{0}}{x_{0}^{3}}\right]  ,\\
q_{a}\left(  x_{0}\right)   &  =-\frac{3}{4}\left[  \frac{\cos x_{0}}{x_{0}%
}+\frac{\sin x_{0}}{x_{0}^{2}}+\frac{\cos x_{0}}{x_{0}^{3}}\right]  ,\\
p_{b}\left(  x_{0}\right)   &  =\frac{3}{4}\left[  \frac{\sin x_{0}}{x_{0}%
}+\frac{3\cos x_{0}}{x_{0}^{2}}-\frac{3\sin x_{0}}{x_{0}^{3}}\right]  ,\\
q_{b}\left(  x_{0}\right)   &  =\frac{3}{4}\left[  -\frac{\cos x_{0}}{x_{0}%
}+\frac{3\sin x_{0}}{x_{0}^{2}}+\frac{3\cos x_{0}}{x_{0}^{3}}\right]  ,
\end{align}
and
\end{subequations}
\begin{equation}
x_{0}=k_{0}R_{21}=\omega_{0}R_{21}/c.
\end{equation}
Note that, for $x_{0}\leq1$, $q_{b}\left(  x_{0}\right)  >0,$ $q_{a}\left(
x_{0}\right)  <0$ and $\omega_{C}>\omega_{A}$ $>\omega_{D}>\omega_{B}$ (see
Fig. \ref{angfig2}).

In this molecular state basis and with our initial conditions, spontaneous
emission and excitation exchange result in a relatively simple decay dynamics
for the molecular state density matrix elements $\rho_{\alpha\alpha^{\prime}%
}\left(  t\right)  $. State $\left\vert E\right\rangle $ undergoes simple
exponential decay at rate $2\gamma_{2}$,
\begin{equation}
\dot{\rho}_{EE}\left(  t\right)  =-2\gamma_{2}\rho_{EE}\left(  t\right)  ,
\end{equation}
population of the ground state is driven only by intermediate state
populations,
\begin{equation}
\dot{\rho}_{GG}\left(  t\right)  =\sum_{\alpha=A}^{D}\gamma_{\alpha}%
\rho_{\alpha\alpha}\left(  t\right)  ,
\end{equation}
and intermediate state populations $\rho_{\alpha\alpha}\left(  t\right)  $
decay at rate $\gamma_{\alpha}$ and are driven by the state $\left\vert
E\right\rangle $ population
\begin{equation}
\dot{\rho}_{\alpha\alpha}\left(  t\right)  =-\gamma_{\alpha}\rho_{\alpha
\alpha}\left(  t\right)  +\frac{\Gamma_{\alpha\alpha}}{2}\rho_{EE}\left(
t\right)  \text{ \ \ \ \ \ \ \ \ \ }\alpha=A,B,C,D,
\end{equation}
with $\Gamma_{AA}=\Gamma_{BB}=\Gamma_{+}$ and $\Gamma_{CC}=\Gamma_{DD}%
=\Gamma_{-}$. In addition, \textit{coherences} $\rho_{AB},$ $\rho_{BA}$,
$\rho_{CD},$ and $\rho_{DC}$ are also driven by the state $\left\vert
E\right\rangle $ population. Details are given in the Appendix.

%EndExpansion

To insure the validity of the RWA and WWA, it is necessary that the level
shifts are much less than $\omega_{0}$, requiring that
\begin{equation}
\frac{\gamma_{2}}{\omega_{0}x_{0}^{3}}\ll1. \label{shifts}%
\end{equation}
Moreover, to be able to neglect retardation in the exchange interaction
between the two atoms as we do, it is necessary that
\begin{subequations}
\label{retard}%
\begin{align}
\gamma_{2}R_{21}/c  &  =\frac{\gamma_{2}x_{0}}{\omega_{0}}\ll1,\\
\left\vert \omega_{\alpha}-\omega_{0}\right\vert R_{21}/c  &  \lesssim
\frac{\gamma_{2}}{\omega_{0}x_{0}^{2}}\ll1;\text{ \ \ \ \ \ }\alpha=A,B,C,D,
\end{align}
and that $t\gg R_{21}/c$.

\section{Source-Field Calculation}

In terms of the electric and magnetic field operators, $\mathbf{E}%
(\mathbf{R},t)\ $and $\mathbf{B}(\mathbf{R},t)$, the expectation value of the
final angular momentum in the field following the decay of the atoms,
$\mathbf{L}_{f}$, is defined as \cite{angmomdef}%
\end{subequations}
\begin{equation}
\mathbf{L}_{f}=\epsilon_{0}\lim_{t\rightarrow\infty}\int_{R_{0}}^{ct}%
R^{2}dR\mathbf{\int}d\Omega\mathbf{\,R}\times\left\langle \mathbf{E}%
(\mathbf{R},t)\times\mathbf{B}(\mathbf{R},t)\right\rangle , \label{1}%
\end{equation}
where $d\Omega$ is an element of solid angle. The radius $R_{0}$ is chosen to
be in the radiation zone of \textit{both} atoms to avoid any issues connected
with initial conditions that involve a sudden excitation of the atom(s)
\cite{class}. Implicit in this assumption is the condition $R_{0}\gg R_{21}$.
A hybrid approach is used, in which the source-field expressions for the field
operators are obtained making neither the RWA nor the WWA, but the subsequent
evaluation of the expectation values of operators are evaluated using both the
RWA and WWA \cite{stokes,pra}.

Since $R_{0}$ is in the radiation zone of both atoms and only field points
with $R>R_{0}$ contribute in Eq. (\ref{1}), you might think it is necessary to
keep only radiation zone terms in the expressions for the field operators.
However, this is not the case. The Poynting vector associated with the
radiation zone fields alone is in the $\mathbf{\hat{R}}$ direction (a caret
over a vector indicates a unit vector) and will lead to a vanishing field
angular momentum. As a consequence it is necessary to keep intermediate zone
terms in the expressions for the fields. This would be true even for a single
atom. \textit{In addition, although }$R_{0}\gg R_{21}$\textit{, one cannot
replace }$\left\vert \mathbf{R}-R_{21}\mathbf{\hat{x}}\right\vert $\textit{ by
}$R$\textit{ in the expressions for the source field operators associated with
atom 2}. The angular momentum in the field is defined relative to the
\textit{origin} of coordinates and the corrections to the field operators
resulting from the fact that atom 2 is displaced from the origin cannot be
ignored when calculating the angular momentum in the field. Such corrections
are not needed when measuring the field energy \cite{ber}.

In RWA and WWA, we can approximate $\mathbf{L}_{f}(t)$ as \cite{pra}
\begin{equation}
\mathbf{L}_{f}=2\epsilon_{0}\lim_{t\rightarrow\infty}\operatorname{Re}%
\int_{R_{0}}^{ct}R^{2}dR\mathbf{\int}d\Omega\mathbf{\,R}\times\left\langle
\mathbf{E}_{s}^{-}(\mathbf{R},t)\times\mathbf{B}_{s}^{+}(\mathbf{R}%
,t)\right\rangle , \label{ang6}%
\end{equation}
in which the fully-retarded source field operators are approximated by
\begin{subequations}
\label{fields}%
\begin{align}
\mathbf{E}_{s}^{-}(\mathbf{R},t)  &  =\frac{\omega_{0}^{2}e^{-ik_{0}%
R}e^{i\omega_{0}t}}{4\pi\epsilon_{0}c^{2}R}\sum_{m=-1}^{1}\left[
\begin{array}
[c]{c}%
\left(  \boldsymbol{\mu}_{-}^{\ast}\left(  m\right)  \cdot\mathbf{\hat{R}%
}\right)  \mathbf{\hat{R}}\left(  \frac{3i}{k_{0}R}-1\right) \\
-\boldsymbol{\mu}_{-}^{\ast}(m)\left(  \frac{i}{k_{0}R}-1\right)
\end{array}
\right]  \sigma_{+}^{(1)}(m,t_{r})\nonumber\\
&  -\frac{\omega_{0}^{2}e^{-ik_{0}R}e^{i\omega_{0}t}}{4\pi\epsilon_{0}c^{2}%
}\sum_{m=-1}^{1}\left[
\begin{array}
[c]{c}%
\frac{\boldsymbol{\mu}_{-}^{\ast}\left(  m\right)  \cdot\left(  \mathbf{R}%
-R_{21}\mathbf{\hat{x}}\right)  }{\left\vert \mathbf{R}-R_{21}\mathbf{\hat{x}%
}\right\vert ^{2}}\left(  \frac{3i}{k_{0}R}-1\right) \\
-\frac{\boldsymbol{\mu}_{-}^{\ast}(m)}{\left\vert \mathbf{R}-R_{21}%
\mathbf{\hat{x}}\right\vert }\left(  \frac{i}{k_{0}R}-1\right)
\end{array}
\right]  \sigma_{+}^{(2)}(m,t_{r})e^{i\mathbf{k}_{0}\cdot\mathbf{R}_{2}},\\
\mathbf{B}_{s}^{+}(\mathbf{R},t)  &  =-\frac{\omega_{0}^{2}e^{ik_{0}%
R}e^{-i\omega_{0}t}}{4\pi\epsilon_{0}c^{3}R}\sum_{m=-1}^{1}\boldsymbol{\mu
}_{-}\left(  m\right)  \times\mathbf{\hat{R}}\left(  \frac{i}{k_{0}%
R}+1\right)  \sigma_{-}^{(1)}(m,t_{r})\nonumber\\
&  -\frac{\omega_{0}^{2}e^{ik_{0}R}e^{-i\omega_{0}t}}{4\pi\epsilon_{0}c^{3}%
}\sum_{m=-1}^{1}\frac{\boldsymbol{\mu}_{-}\left(  m\right)  \times\left(
\mathbf{R}-R_{21}\mathbf{\hat{x}}\right)  }{\left\vert \mathbf{R}%
-R_{21}\mathbf{\hat{x}}\right\vert ^{2}}\left(  \frac{i}{k_{0}R}+1\right)
\sigma_{-}^{(2)}(m,t_{r})e^{-i\mathbf{k}_{0}\cdot\mathbf{R}_{2}},
\end{align}
where%
\end{subequations}
\begin{equation}
\boldsymbol{\mu}_{-}\left(  m\right)  =\mu_{12}\left[  \frac{\left(
\delta_{m,1}-\delta_{m,-1}\right)  \mathbf{\hat{x}}+i\left(  \delta
_{m,1}+\delta_{m,-1}\right)  \mathbf{\hat{y}}}{\sqrt{6}}\mathbf{+}\frac
{\delta_{m,0}\mathbf{\hat{z}}}{\sqrt{3}}\right]  , \label{fp2}%
\end{equation}
$\mathbf{k}_{0}=k_{0}\mathbf{\hat{R}}$, $\sigma_{+}^{(j)}(m,t_{r})$ is a
raising operator for atom $j$ connecting the ground state to the $m$ excited
state, $\sigma_{-}^{(j)}(m,t_{r})$ is a lowering operator for atom $j$
connecting the $m$ excited state to the ground state, and $t_{r}=t-R/c$ is the
retarded time. Only radiation and intermediate zone field contributions have
been retained. For our initial conditions, only the $x$ and $y$ components of
the dipole moment operators enter the calculation. The raising and lowering
operators are written in an interaction representation, $\sigma_{\pm}%
^{total}(m,t)=\sigma_{\pm}(m,t)e^{\pm i\omega_{0}t}$.

When evaluating the expectation values of the product of operators given in
Eq. (\ref{ang6}) using the field operators given in Eqs. (\ref{fields}), two
types of terms are encountered. The first are "population" terms of the form
$\left\langle \sigma_{+}^{(j)}(m,t)\sigma_{-}^{(j)}(m^{\prime},t)\right\rangle
=\left\langle \sigma^{(j)}(m,m^{\prime},t)\right\rangle $ involving single
atom operators, and the second are "coherence" terms of the form $\left\langle
\sigma_{+}^{(j)}(m,t)\sigma_{-}^{(j^{\prime})}(m^{\prime},t)\right\rangle $
with\ $j\neq j^{\prime}$, involving products of operators of different atoms.
Expressions for these quantities are given in the Appendix in terms of
molecular state density matrix elements.

After considerable algebra, we find that, in the limit that $R\rightarrow
\infty$ (that is, in the radiation zone) \cite{math},
\begin{align}
\mathbf{L}_{f}  &  =\hbar\mathbf{\hat{z}}\left[  S_{AB}+S_{BA}+S_{CD}%
+S_{DC}+\rho_{EE}(0)\right] \nonumber\\
&  +\hbar p_{a}\mathbf{\hat{z}}\left(  S_{AB}+S_{BA}-S_{CD}-S_{DC}\right)
\nonumber\\
&  +\hbar p_{b}\mathbf{\hat{z}}\left(  S_{AC}+S_{CA}-S_{BD}-S_{DB}\right)  ,
\label{elf}%
\end{align}
where the $x_{0}$ dependence of $p_{a}$ and $p_{b}$ has been suppressed,
\begin{equation}
S_{\alpha\alpha^{\prime}}=\frac{\gamma_{2}}{c}\lim_{t\rightarrow\infty}%
\int_{R_{0}}^{ct}\rho_{\alpha\alpha^{\prime}}\left(  t-R/c\right)  dR,
\label{es}%
\end{equation}
and values of density matrix elements $\rho_{\alpha\alpha^{\prime}}\left(
t\right)  $ are given in Eqs.(\ref{rhode}) and (\ref{rhose}) of the Appendix.

For the DE initial condition, we can use the values of $S_{\alpha
\alpha^{\prime}}^{DE}$ given in Eq. (\ref{esde}) of the Appendix to arrive at
\begin{equation}
\mathbf{L}_{f}^{DE}=\hbar\mathbf{\hat{z}}\left[  1+\frac{\left(
1+p_{a}\right)  ^{3}}{2\left[  \left(  1+p_{a}\right)  ^{2}+q_{b}^{2}\right]
}+\frac{\left(  1-p_{a}\right)  ^{3}}{2\left[  \left(  1-p_{a}\right)
^{2}+q_{b}^{2}\right]  }\right]  . \label{elfDE}%
\end{equation}
If $x_{0}\gg1$, $p_{a},q_{b}\sim0$ and $\mathbf{L}_{f}=2\hbar\mathbf{\hat{z}}%
$. As expected, the total initial angular momentum of the atoms, $\left\langle
\mathbf{L}_{atoms}^{DE}(t=0)\right\rangle =2\hbar\mathbf{\hat{z}}$, has been
transferred to the field since the atom-atom interactions are negligible in
this limit. On the other hand, for $x_{0}\ll1$, $p_{a}\sim1,$ $q_{b}\sim
\infty$, and $\mathbf{L}_{f}=\hbar\mathbf{\hat{z}}$. Owing to the rapid
exchange between the $m=1$ and $m=-1$ sublevels following the emission of the
first photon (which transfers $\hbar$ of angular momentum to the field), there
is no further transfer of angular momentum from the atoms to the field.

For the SE initial condition, we can use the values of $S_{\alpha
\alpha^{\prime}}^{SE}$ given in Eq. (\ref{esse}) of the Appendix to obtain
\begin{align}
\mathbf{L}_{f}^{SE}  &  =\hbar\mathbf{\hat{z}}\left[  \frac{\left(
1+p_{a}\right)  ^{2}}{2\left[  \left(  1+p_{a}\right)  ^{2}+q_{b}^{2}\right]
}+\frac{\left(  1-p_{a}\right)  ^{2}}{2\left[  \left(  1-p_{a}\right)
^{2}+q_{b}^{2}\right]  }\right] \nonumber\\
&  +\hbar p_{b}\mathbf{\hat{z}}\left[  \frac{\left(  1+p_{b}\right)
}{2\left[  \left(  1+p_{b}\right)  ^{2}+q_{a}^{2}\right]  }-\frac{\left(
1-p_{b}\right)  }{2\left[  \left(  1-p_{b}\right)  ^{2}+q_{a}^{2}\right]
}\right] . \label{elfSE}%
\end{align}
If $x_{0}\gg1$, $p_{a},p_{b},q_{a},q_{b}\sim0$ and $\mathbf{L}_{f}%
=\hbar\mathbf{\hat{z}}$. As expected, the total initial angular momentum of
the atoms, $\left\langle \mathbf{L}_{atoms}^{SE}(t=0)\right\rangle
=\hbar\mathbf{\hat{z}}$ has been transferred to the field since the atom-atom
interactions are negligible in this limit. On the other hand, for $x_{0}\ll1$,
$p_{a}\sim1,$ $p_{b}\sim0,$ $q_{a}\sim-\infty$, $q_{b}\sim\infty$, and
$\mathbf{L}_{f}=0\mathbf{\hat{z}}$. Owing to the rapid exchange between the
$m=1$ and $m=-1$ sublevels, there is no transfer of angular momentum from the
atoms to the field.

\begin{figure}[ptb]
\centering
\includegraphics[width=\textwidth,height=\textheight,keepaspectratio
]{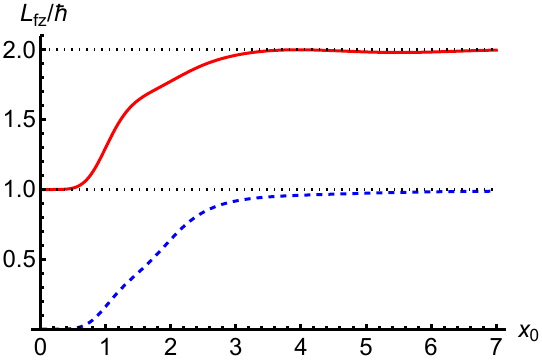}\caption{Plots of the $z-$component of the final field angular
momentum, $L_{fz}$, (in units of $\hbar$) as a function of $x_{0}=k_{0}R_{21}$
for the DE and SE initial conditions. For the DE initial condition the initial
angular momentum in the atoms is $\left\langle L_{atom}^{DE}(0)\right\rangle
=2\hbar$ and for the SE initial condition the initial angular momentum in the
atoms is $\left\langle L_{atom}^{SE}(0)\right\rangle =\hbar$. }%
\label{angfig3}%
\end{figure}

In Fig. \ref{angfig3}, the $z-$ component of the final field angular momentum
(in units of $\hbar$) is plotted as a function of $x_{0}$ for the DE (solid,
red curve) and SE (dashed, blue curve) initial conditions. In general, for
arbitrary $x_{0}$, not all of the initial angular momentum of the atoms is
transferred to the fields. To conserve angular momentum, some angular momentum
must be converted into orbital angular momentum of the two-atom system. We now
calculate this quantity to see if total angular momentum is conserved.

%EndExpansion

\bigskip

\section{\bigskip Orbital Angular Momentum of the 2-Atom System}

In RWA, the expectation value of the force on atom 2 exerted by the source
field of atom 1, $\mathbf{F}_{21}(t)$, can be taken as%
\begin{align}
\mathbf{F}_{21}(\mathbf{R}_{2},t)  &  =2\operatorname{Re}\left[
e^{i\omega_{0}t}\sum_{m=-1}^{1}\left\langle \sigma_{+}^{(2)}(m,t)\left(
\boldsymbol{\mu}_{-}^{\ast}\left(  m\right)  \cdot\boldsymbol{\nabla}\right)
\mathbf{E}_{s}^{+}(\mathbf{R},t)\right.  \right. \nonumber\\
&  +\left.  \left.  i\omega_{0}\sigma_{+}^{(2)}(m,t)\boldsymbol{\mu}_{-}%
^{\ast}\left(  m\right)  \times\mathbf{B}_{s}^{+}(\mathbf{R},t)\right\rangle
_{\mathbf{R}=\mathbf{R}_{2}}\right]  , \label{force}%
\end{align}

where%

\begin{subequations}
\label{orbf}%
\begin{align}
\mathbf{E}_{s}^{+}(\mathbf{R},t)  &  =\frac{\omega_{0}^{2}e^{ik_{0}%
R}e^{-i\omega_{0}t}}{4\pi\epsilon_{0}c^{2}R}\sum_{m=-1}^{1}\left[
\begin{array}
[c]{c}%
\left(  \boldsymbol{\mu}_{-}\left(  m\right)  \cdot\mathbf{\hat{R}}\right)
\mathbf{\hat{R}}\left(  \frac{3}{\left(  k_{0}R\right)  ^{2}}-\frac{3i}%
{k_{0}R}-1\right) \\
-\boldsymbol{\mu}_{-}(m)\left(  \frac{1}{\left(  k_{0}R\right)  ^{2}}-\frac
{i}{k_{0}R}-1\right)
\end{array}
\right]  \sigma_{-}^{(1)}(m,t),\\
\mathbf{B}_{s}^{+}(\mathbf{R},t)  &  =-\frac{\omega_{0}^{2}e^{ik_{0}%
R}e^{-i\omega_{0}t}}{4\pi\epsilon_{0}c^{3}R}\sum_{m=-1}^{1}\boldsymbol{\mu
}_{-}\left(  m\right)  \times\mathbf{\hat{R}}\left[  \frac{i}{k_{0}%
R}+1\right]  \sigma_{-}^{(1)}(m,t).
\end{align}
The electric field expression now includes a near zone contribution since the
field is evaluated at the position of atom 2, which may lie in the near zone.
Moreover, the interaction representation raising and lowering operators are
evaluated at time $t$ rather than $t_{r}$, assuming inequality (\ref{retard})
holds. \ Equation (\ref{force}) is the quantum analogue of the classical
expression of the force $\mathbf{F}_{class}(\mathbf{R},t)$ on an oscillating
dipole having dipole moment $\mathbf{p}(t)$ that is located at position
$\mathbf{R}$, namely \cite{torque}
\end{subequations}
\begin{equation}
\mathbf{F}_{class}(\mathbf{R},t)=\left(  \mathbf{p}(t)\cdot\boldsymbol{\nabla
}\right)  \mathbf{E}(\mathbf{R},t)+\mathbf{\dot{p}}(t)\times\mathbf{B}%
(\mathbf{R},t).
\end{equation}

To calculate the final orbital angular momentum of the two-atom system, we
make the simplifying assumption that the atoms are infinitely massive, in
order to neglect any motion of the atoms. In that limit, the final orbital
angular momentum, $\mathbf{L}_{orb},$ is given by%
\begin{equation}
\mathbf{L}_{orb}=\mathbf{R}_{2}\mathbf{\times}\int_{0}^{\infty}dt\mathbf{F}%
_{21}(\mathbf{R}_{2},t) \label{elorb}%
\end{equation}
In contrast to the calculation of the field angular momentum, there is only a
"coherence" contribution to the orbital angular momentum. From Eqs.
(\ref{force}) - (\ref{elorb}) and the equations in the Appendix, we find
\cite{math}%
\begin{align}
\mathbf{L}_{orb}  &  =i\hbar q_{b}\mathbf{\hat{z}}\left(  S_{AB}-S_{BA}%
+S_{CD}-S_{DC}\right) \nonumber\\
&  -\hbar p_{b}\mathbf{\hat{z}}\left(  S_{AC}+S_{CA}-S_{BD}-S_{DB}\right)  .
\end{align}

For the DE initial condition,%
\begin{equation}
\mathbf{L}_{orb}^{DE}=\hbar\mathbf{\hat{z}}\left[  \frac{\left(
1+p_{a}\right)  q_{b}^{2}}{2\left[  \left(  1+p_{a}\right)  ^{2}+q_{b}%
^{2}\right]  }+\frac{\left(  1-p_{a}\right)  q_{b}^{2}}{2\left[  \left(
1-p_{a}\right)  ^{2}+q_{b}^{2}\right]  }\right]
\end{equation}
For the SE initial condition,%
\begin{align}
\mathbf{L}_{orb}^{SE}  &  =\hbar\mathbf{\hat{z}}\left[  \frac{q_{b}^{2}%
}{2\left[  \left(  1+p_{a}\right)  ^{2}+q_{b}^{2}\right]  }+\frac{q_{b}^{2}%
}{2\left[  \left(  1-p_{a}\right)  ^{2}+q_{b}^{2}\right]  }\right] \nonumber\\
&  -\hbar p_{b}\mathbf{\hat{z}}\left[  \frac{\left(  1+p_{b}\right)
}{2\left[  \left(  1+p_{b}\right)  ^{2}+q_{a}^{2}\right]  }-\frac{\left(
1-p_{b}\right)  }{2\left[  \left(  1-p_{b}\right)  ^{2}+q_{a}^{2}\right]
}\right]
\end{align}

Combining these results with Eqs. (\ref{elfDE}) and (\ref{elfSE}), we find
that the sum of the total final angular momentum of the field and the orbital
angular momentum of the two-atom system is equal to the original internal
angular momentum of the atoms,%
\begin{subequations}
\begin{align}
\mathbf{L}_{total}^{DE}  &  =\mathbf{L}_{f}^{DE}+\mathbf{L}_{orb}^{DE}%
=2\hbar\mathbf{\hat{z},}\\
\mathbf{L}_{total}^{SE}  &  =\mathbf{L}_{f}^{SE}+\mathbf{L}_{orb}^{SE}%
=\hbar\mathbf{\hat{z}.}%
\end{align}
Both the DE and SE results are consistent with conservation of total angular momentum.

\section{Schr\"{o}dinger Picture Approach}

An alternative method for calculating the angular momentum in the field was
developed in Ref. \cite{jmo}. In that paper, a Schr\"{o}dinger picture
approach was taken to evaluate the angular momentum in the field radiated by a
single atom. An advantage of the Schr\"{o}dinger picture approach is that it
can be used to calculate the field angular momentum at any time. The results
in Ref. \cite{jmo} can be generalized easily to the case of two atoms for the
SE initial condition in which only one of the atoms is in its excited state.
For simplicity, that is the only initial condition considered in this section.
In RWA we can write the state vector for times $t>0$ in an interaction
representation as
\end{subequations}
\begin{equation}
\left\vert \psi(t)\right\rangle =\sum_{\alpha=A}^{D}b_{\alpha;0}%
(t)e^{-i\omega_{0}t}\left\vert \alpha;0\right\rangle +\int d\mathbf{k}%
\sum_{\lambda}b_{G}^{(\lambda)}(\mathbf{k},t)e^{-i\omega_{k}t}\left\vert
G;\mathbf{k}_{\lambda}\right\rangle , \label{stvec}%
\end{equation}
where $b_{\alpha;0}(t)$ is the state amplitude for the atoms to be in state
$\left\vert \alpha\right\rangle $ and the field to be in the vacuum state at
time $t$ and $b_{G}^{(\lambda)}(\mathbf{k},t)$ is the state amplitude for the
atoms to be in the ground state $\left\vert G\right\rangle $ and the field to
be in state $\left\vert \mathbf{k}_{\lambda}\right\rangle $ at time $t$. The
SE initial condition in the molecular basis is%
\begin{equation}
\left\vert \psi(0)\right\rangle =\frac{1}{2}\sum_{\alpha=A}^{D}\left\vert
\alpha;0\right\rangle ; \label{psi0}%
\end{equation}
that is
\begin{equation}
b_{\alpha;0}(0)=1/2;\text{ \ \ \ \ \ \ \ \ }\alpha=A,B,C,D.
\end{equation}

In WWA, the excited-state amplitudes undergo exponential decay given by
\begin{equation}
b_{\alpha;0}(t)=\frac{1}{2}e^{-\gamma_{\alpha}t/2}e^{-i(\omega_{\alpha}%
-\omega_{0})t}, \label{expdec}%
\end{equation}
where $\gamma_{\alpha}$ and $\omega_{\alpha}$ are given in Eqs. (\ref{decpar})
and (\ref{freqpar}), respectively. The time evolution equation for
$b_{G}^{(\lambda)}(\mathbf{k},t)$ is then
\begin{equation}
\dot{b}_{G}^{(\lambda)}(\mathbf{k},t)=\frac{1}{2i\hbar}\sum_{\alpha=A}%
^{D}\left\langle G;\mathbf{k}_{\lambda}\right\vert V\left\vert \alpha
;0\right\rangle e^{i\left(  \omega_{k}-\omega_{\alpha}\right)  t}%
e^{-\gamma_{\alpha}t/2}, \label{amp3}%
\end{equation}
where $V$ is the interaction Hamiltonian given by Eq. (\ref{hrwa}) in the Appendix.

Equation (\ref{amp3}) can be integrated to obtain
\begin{align}
b_{G}^{(\lambda)}(\mathbf{k},t)  &  =\frac{\mu_{12}}{2\left(  2\pi\right)
^{3/2}}\left(  \frac{\omega_{k}}{2\hbar\epsilon_{0}}\right)  ^{1/2}%
\sum_{\alpha=A}^{D}M_{\alpha}^{\left(  \lambda\right)  }\left(  \mathbf{k}%
\right) \nonumber\\
&  \times\frac{1-e^{-\gamma_{\alpha}t/2}e^{i\left(  \omega_{k}-\omega_{\alpha
}\right)  t}}{\gamma_{\alpha}/2-i\left(  \omega_{k}-\omega_{\alpha}\right)  },
\label{examp}%
\end{align}
where the $M_{\alpha}^{\left(  \lambda\right)  }\left(  \mathbf{k}\right)  $
matrix elements are given in Eq. (\ref{matel}) of the Appendix. We are
interested in the final angular momentum in the field, so we take the limit
that $t\rightarrow\infty$ and write
\begin{equation}
b_{G}^{(\lambda)}(\mathbf{k})\equiv b_{G}^{(\lambda)}(\mathbf{k},\infty
)=\frac{\mu_{12}}{2\left(  2\pi\right)  ^{3/2}}\left(  \frac{\omega_{k}%
}{2\hbar\epsilon_{0}}\right)  ^{1/2}\sum_{\alpha=A}^{D}\frac{M_{\alpha
}^{\left(  \lambda\right)  }\left(  \mathbf{k}\right)  }{\gamma_{\alpha
}/2-i\left(  \omega_{k}-\omega_{\alpha}\right)  }.
\end{equation}

In Ref. \cite{jmo} , the total angular momentum operator of the field in RWA
is expressed in terms of creation and annihilation operators as%
\begin{align}
\mathbf{L}_{f}  &  =-\hbar c\frac{1}{2\left(  2\pi\right)  ^{3}}\int
d\mathbf{k}\int d\mathbf{k}^{\prime}\sum_{\lambda,\lambda^{\prime}}a^{\dag
}\left(  \mathbf{k}_{\lambda^{\prime}}^{\prime}\right)  a\left(
\mathbf{k}_{\lambda}\right)  \sqrt{\omega_{k}\omega_{k^{\prime}}}\nonumber\\
&  \times\left[  \boldsymbol{\epsilon}_{\mathbf{k}}^{(\lambda)}\times\left(
\mathbf{\hat{k}}^{\prime}\mathbf{\times}\boldsymbol{\epsilon}_{\mathbf{k}%
^{\prime}}^{(\lambda^{\prime})}\right)  \right]  \times\int d\mathbf{R\,R}%
e^{i\left(  \mathbf{k}-\mathbf{k}^{\prime}\right)  \cdot\mathbf{R}}+adjoint.
\label{poy}%
\end{align}
This expression is formally equivalent to Eq. (\ref{ang6}), so we would expect
it to reproduce the source-field result for the final angular momentum in the
field. In WWA, the expectation value of $\mathbf{L}_{f}$ can be written as
\cite{jmo}
\begin{align}
\left\langle \mathbf{L}_{f}(t)\right\rangle  &  =\left(  i\frac{\hbar
\omega_{0}}{2c}\sum_{\lambda,\lambda^{\prime}}\int d\mathbf{k\,}%
b_{G}^{(\lambda^{\prime})}(\mathbf{k},t)^{\ast}\left[  \left(  \mathbf{\hat
{k}\times}\boldsymbol{\epsilon}_{\mathbf{k}}^{(\lambda^{\prime})}\right)
\cdot\mathbf{\nabla}_{\mathbf{k}}\right]  \left[  b_{G}^{(\lambda)}%
(\mathbf{k},t)\boldsymbol{\epsilon}_{\mathbf{k}}^{(\lambda)}\right]
+c.c.\right) \nonumber\\
&  +\left(  -i\frac{\hbar\omega_{0}}{2c}\sum_{\lambda,\lambda^{\prime}}\int
d\mathbf{k\,}b_{G}^{(\lambda^{\prime})}(\mathbf{k},t)^{\ast}\left(
\mathbf{\hat{k}\times}\boldsymbol{\epsilon}_{\mathbf{k}}^{(\lambda^{\prime}%
)}\right)  \mathbf{\nabla}_{\mathbf{k}}\cdot\left[  b_{G}^{(\lambda
)}(\mathbf{k},t)\boldsymbol{\epsilon}_{\mathbf{k}}^{(\lambda)}\right]
+c.c.\right)  . \label{36}%
\end{align}
where $c.c.$ stands for "complex conjugate." For a single atom at the origin,
the first term vanishes on integration over solid angle (in $k-$space);
however, this is no longer true for our 2-atom system. Following the decay of
the atoms, the $z-$component of field angular momentum obtained from Eq.
(\ref{36}), $\left\langle L_{f}(\infty)\right\rangle _{z}$, reduces to%
\begin{equation}
\left\langle L_{f}(\infty)\right\rangle _{z}=-i\frac{\hbar}{2}\int
d\mathbf{k}\left[  b_{G}^{(\theta)}(\mathbf{k})^{\ast}\frac{\partial
b_{G}^{(\theta)}(\mathbf{k})}{\partial\phi_{k}}+b_{G}^{(\phi)}(\mathbf{k}%
)^{\ast}\frac{\partial b_{G}^{(\phi)}(\mathbf{k})}{\partial\phi_{k}}\right]
+c.c. \label{angtot}%
\end{equation}

Using Eqs. (\ref{angtot}), (\ref{amp3}), and (\ref{matel}), we can carry out
the integrations in WWA assuming that inequalities (\ref{shifts}) and
(\ref{retard}) are satisfied to obtain \cite{math}%
\begin{align}
\mathbf{L}_{f}^{SE}  &  =\hbar\mathbf{\hat{z}}\left[  \frac{\left(
1+p_{a}\right)  ^{2}}{2\left[  \left(  1+p_{a}\right)  ^{2}+q_{b}^{2}\right]
}+\frac{\left(  1-p_{a}\right)  ^{2}}{2\left[  \left(  1-p_{a}\right)
^{2}+q_{b}^{2}\right]  }\right] \nonumber\\
&  +\hbar p_{b}\mathbf{\hat{z}}\left[  \frac{\left(  1+p_{b}\right)
}{2\left[  \left(  1+p_{b}\right)  ^{2}+q_{a}^{2}\right]  }-\frac{\left(
1-p_{b}\right)  }{2\left[  \left(  1-p_{b}\right)  ^{2}+q_{a}^{2}\right]
}\right]  ,
\end{align}
in agreement with the source field result given in Eq. (\ref{elfSE}).

\section{Discussion}

Two-atom Dicke superradiance is a problem of fundamental importance in quantum
optics. Many aspects of the problem have been studied, starting with
Lehmberg's detailed solution of the dynamics of superradiant decay from two,
two-level atoms \cite{leh}. When one allows for an excited state having Zeeman
degeneracy, a whole new set of features can arise. In this paper we have
studied how some initial internal angular momentum of the combined two-atom
system is transferred to the field radiated by the atoms. We used two,
complementary approaches to obtain the final field angular momentum, a
source-field approach and a Schr\"{o}dinger equation approach. Both approaches
gave identical answers, but it was necessary to include terms in both
calculations that, at first glance, might not be expected to contribute. We
have shown that the final angular momentum in the field is not necessarily
equal to the initial angular momentum in the two-atom system, owing to an
exchange between a state in which (atom 1 is in its $m=1$ excited state and
atom 2 in its ground state) to one in which (atom 2 is in its $m=-1$ excited
state and atom 1 in its ground state). The missing angular momentum appears as
orbital angular momentum of the two atoms. To facilitate the calculation, it
has been assumed that the atoms are infinitely massive, allowing us to neglect
any motion of the atoms resulting from the force between the atoms. This is
not a particularly good approximation if the atoms are separated by less than
a wavelength \cite{ber3}. In some sense, the fact that we obtained results
consistent with conservation of angular momentum \ is somewhat remarkable. We
used source-field expressions which were derived from a Hamiltonian without
making the RWA and then evaluated the needed expectation values using the RWA
and WWA. The amplitude calculations were made by starting with the RWA
Hamiltonian and then imposing the WWA. Despite all these approximations, it
was still found that angular momentum was conserved. The Schr\"{o}dinger
equation approach is especially well-suited to a study of the time-dependence
of the field angular momentum \cite{jmo}, even though we have concentrated
only on the final angular momentum in the field in this paper.

As was noted in the Introduction, we considered only two atoms and a $J=1$ to
$J=0$ transition in order to illustrate the underlying physics. The extension
to decay from atoms having an excited state with with total angular momentum
$F=F_{2}$ to a ground state having total angular momentum $F=F_{1}=$ $F_{2}-1$
complicates the calculation, but adds little new physics. For example if the
two atoms on the $x-$axis are prepared in the $m=2F_{2}+1$ excited state
sublevels, one would find that excitation exchange results in $4\left(
F_{2}\right)  ^{2}$ intermediate molecular states for integral $F_{2}$ and
$\left[  4\left(  F_{2}\right)  ^{2}-1\right]  $ for 1/2 integral $F_{2}$,
necessitating a numerical rather than analytic solution to the problem for
$F_{2}>1$. The atoms start with $2\left(  2F_{2}+1\right)  \hbar$ of internal
state angular momentum and the fields radiated during spontaneous decay
contain at most $2\hbar$ of angular momentum after the atoms have decayed. Any
difference in angular momentum between the initial internal angular momentum
in the atoms and the final angular momentum of the fields appears as orbital
angular momentum of the atoms and internal ground state angular momentum of
the atoms \cite{jmo}. Even for a $J=1$ to $J=0$ transition, the extension of
the calculation to $N$ atoms greatly complicates things since the number of
relevant intermediate molecular states scales as $\sum_{n=1}^{N-1}C_{n}%
^{N}3^{N-n}$ for atoms at arbitrary positions prepared in their $m=1$ excited
state sublevels, where $C_{n}^{N}$ is a binomial coefficient. The initial
internal angular momentum of the atoms it converted to the angular momentum of
the fields plus the relative orbital angular momentum of all atom pairs. In
other words, generalization of the calculation to transitions between states
having arbitrary angular momentum and to systems of $N$ atoms greatly
complicates the calculation, but does not result in any changes to the
underlying physics.

Admittedly, experimental tests of the theory presented in this paper would
prove challenging, although advances in atom arrays and ultrafast optics might
provide a pathway. One could envision the excitation of two closely spaced
atoms in the array using an ultrafast pulse, whose subsequent radiation is
then probed by other atoms in the same or a neighboring array. The detector
atoms could serve as a measure of the angular momentum in the field. A $J=1$
to $J=0$ transition could be realized using $^{88}$Sr atoms.

\bigskip

PRB would like to acknowledge helpful discussions with D. G. Steel and P. W. Milonni.

\bigskip

This research is supported by the Air Force Office of Scientific Research and
the National Science Foundation.

\bigskip\appendix

\section{Calculation Details}

In this Appendix some of the details of the calculations are given
\cite{math}. In dipole approximation and RWA, the interaction Hamiltonian $V$
for the atom-field system can be written as%
\begin{equation}
V=-\sum_{j=1}^{2}\left[  \boldsymbol{\mu}_{+}^{(j)}\cdot\mathbf{E}_{+}\left(
\mathbf{R}_{j}\right)  +\mathbf{E}_{+}\left(  \mathbf{R}_{j}\right)
\cdot\boldsymbol{\mu}_{-}^{(j)}\right]  , \label{hrwa}%
\end{equation}
where
\begin{align}
\boldsymbol{\mu}_{+}^{(j)}  &  =\sum_{m=-1}^{1}\left[  \boldsymbol{\mu}%
_{-}\left(  m\right)  \right]  ^{\ast}\left(  \left\vert m\right\rangle
\left\langle g\right\vert \right)  _{j},\\
\boldsymbol{\mu}_{-}^{(j)}  &  =\sum_{m=-1}^{1}\boldsymbol{\mu}_{-}\left(
m\right)  \left(  \left\vert g\right\rangle \left\langle m\right\vert \right)
_{j},
\end{align}
$\left(  \left\vert m\right\rangle \left\langle g\right\vert \right)  _{j}$ is
a raising operator for atom $j$, $\left(  \left\vert g\right\rangle
\left\langle m\right\vert \right)  _{j}$ is a lowering operator for atom $j$,
$\boldsymbol{\mu}_{-}\left(  m\right)  $ is defined by Eq. (\ref{fp2}),
\begin{equation}
\mathbf{E}_{+}\left(  \mathbf{R}\right)  =\frac{i}{\left(  2\pi\right)
^{3/2}}\sum_{\lambda}\int d\mathbf{k}\left(  \frac{\hbar\omega_{k}}%
{2\epsilon_{0}}\right)  ^{1/2}a\left(  \mathbf{k}_{\lambda}\right)
\boldsymbol{\epsilon}_{\mathbf{k}}^{(\lambda)}e^{i\mathbf{k\cdot R}}=\left[
\mathbf{E}_{-}\left(  \mathbf{R}\right)  \right]  ^{\dag}%
\end{equation}
$a\left(  \mathbf{k}_{\lambda}\right)  $ and $a^{\dag}\left(  \mathbf{k}%
_{\lambda}\right)  $ are annihilation and creation operators for a field mode
having wave vector%
\begin{equation}
\mathbf{k}=k\left(  \sin\theta_{k}\cos\phi_{k}\mathbf{\hat{x}}+\sin\theta
_{k}\sin\phi_{k}\mathbf{\hat{y}+}\cos\theta_{k}\mathbf{\hat{z}}\right)  ,
\end{equation}
frequency $\omega_{k}=kc,$ and polarization $\boldsymbol{\epsilon}%
_{\mathbf{k}}^{(\lambda)},$ with
\begin{subequations}
\label{ep}%
\begin{align}
\boldsymbol{\epsilon}_{\mathbf{k}}^{(1)}  &  =\boldsymbol{\epsilon
}_{\mathbf{k}}^{(\theta)}=\cos\theta_{k}\cos\phi_{k}\mathbf{\hat{x}}%
+\cos\theta_{k}\sin\phi_{k}\mathbf{\hat{y}}-\sin\theta_{k}\mathbf{\hat{z},}\\
\boldsymbol{\epsilon}_{\mathbf{k}}^{(2)}  &  =\boldsymbol{\epsilon
}_{\mathbf{k}}^{(\phi)}=-\sin\phi_{k}\mathbf{\hat{x}}+\cos\phi_{k}%
\mathbf{\hat{y}.}%
\end{align}

Using this interaction Hamiltonian and starting from an initial state in which
both atoms share at most one excitation of an $m=1$ excited state sublevel, we
can write the state vector (in an interaction representation) as
\end{subequations}
\begin{equation}
\left\vert \psi(t)\right\rangle =e^{-i\omega_{0}t}\sum_{m=\pm1}\left[
b_{mg}(t)\left\vert 1g\right\rangle +b_{gm}(t)\left\vert g1\right\rangle
\right]  +\sum_{\lambda}\int d\mathbf{k}b_{gg}^{(\lambda)}(\mathbf{k}%
,t)e^{-i\omega_{k}t}\left\vert gg;\mathbf{k}_{\lambda}\right\rangle ,
\end{equation}
where $b_{mg}(t)$ is the state amplitude for atom 1 to be in its $m$ excited
state sublevel, atom 2 to be in its ground state, and the field to be in the
vacuum state at time $t$, $b_{gm}(t)$ is the state amplitude for atom 2 to be
in its $m$ excited state sublevel, atom 1 to be in its ground state, and the
field to be in the vacuum state at time $t$, and $b_{gg}^{(\lambda
)}(\mathbf{k},t)$ is the state amplitude for the both atoms to be in their
ground states and the field to be in state $\left\vert \mathbf{k}_{\lambda
}\right\rangle $ at time $t$. \ Eliminating the state amplitude $b_{gg}%
^{(\lambda)}(\mathbf{k},t)$ using standard methods, we can obtain the coupled
equations for the excited state amplitudes in WWA as \cite{fu}
\begin{subequations}
\label{ev}%
\begin{align}
\dot{b}_{1g}  &  =-\gamma b_{1g}-\gamma\xi_{a}b_{g1}-\gamma\xi_{b}b_{g,-1},\\
\dot{b}_{g1}  &  =-\gamma b_{g1}-\gamma\xi_{a}b_{1g}-\gamma\xi_{b}b_{-1g},\\
\dot{b}_{-1g}  &  =-\gamma b_{-1g}-\gamma\xi_{a}b_{g,-1}-\gamma\xi_{b}%
b_{g1},\\
\dot{b}_{g,-1}  &  =-\gamma b_{g,-1}-\gamma\xi_{a}b_{-1g}-\gamma\xi_{b}b_{1g},
\end{align}
where
\end{subequations}
\begin{align}
\xi_{a}  &  =p_{a}\left(  k_{0}R_{21}\right)  +iq_{a}\left(  k_{0}%
R_{21}\right)  ,\\
\xi_{b}  &  =p_{b}\left(  k_{0}R_{21}\right)  +iq_{b}\left(  k_{0}%
R_{21}\right)  .
\end{align}
and $p_{a},q_{a},p_{b},q_{b}$ are given in Eqs. (\ref{pq}).

By diagonalizing Eqs. (\ref{ev}), we arrive at the molecular basis defined in
Eqs. (\ref{molbas}). In this basis, the density matrix elements evolve as
\cite{ber}%
\begin{subequations}
\begin{align}
\dot{\rho}_{GG}\left(  t\right)   &  =\gamma_{A}\rho_{AA}\left(  t\right)
+\gamma_{B}\rho_{BB}\left(  t\right)  +\gamma_{C}\rho_{CC}\left(  t\right)
+\gamma_{D}\rho_{DD}\left(  t\right)  ,\\
\dot{\rho}_{\alpha\alpha^{\prime}}\left(  t\right)   &  =-\left(
\gamma_{\alpha\alpha^{\prime}}+i\omega_{\alpha\alpha^{\prime}}\right)
\rho_{\alpha\alpha^{\prime}}\left(  t\right)  +\frac{\Gamma_{\alpha
\alpha^{\prime}}}{2}\rho_{EE}\left(  t\right)  ,\text{ \ \ \ }\alpha
,\alpha^{\prime}=A,B,C,D\\
\dot{\rho}_{EE}\left(  t\right)   &  =-2\gamma_{2}\rho_{EE}\left(  t\right)  ,
\end{align}
where
\end{subequations}
\begin{subequations}
\begin{align}
\Gamma_{AA}  &  =\Gamma_{BB}=\Gamma_{AB}=\Gamma_{BA}=\Gamma_{+}=\gamma
_{2}\left(  1+p_{a}\right)  \text{,}\\
\text{ \ }\Gamma_{CC}  &  =\Gamma_{DD}=\Gamma_{CD}=\Gamma_{DC}=\Gamma
_{-}=\gamma_{2}\left(  1-p_{a}\right)  ,\\
\Gamma_{AC}  &  =\Gamma_{CA}=\Gamma_{BD}=\Gamma_{DB}=\Gamma_{AD}=\Gamma
_{DA}=\Gamma_{BC}=\Gamma_{CB}=0,\\
\gamma_{\alpha\alpha^{\prime}}  &  =\frac{\gamma_{\alpha}+\gamma
_{\alpha^{\prime}}}{2}\text{; \ \ \ \ \ }\omega_{\alpha\alpha^{\prime}}%
=\omega_{\alpha}-\omega_{\alpha^{\prime}}\text{; \ \ }\alpha,\alpha^{\prime
}=A,B,C,D,
\end{align}
and values of $\gamma_{\alpha}$ and $\omega_{\alpha}$ are given in Eqs.
(\ref{decpar}) and (\ref{freqpar}), respectively.

The solution for the DE initial condition,%
\end{subequations}
\begin{equation}
\rho_{EE}\left(  0\right)  =1\text{; \ \ \ }\rho_{GG}\left(  0\right)
=\rho_{\alpha\alpha^{\prime}}\left(  0\right)  =0\text{ \ \ \ for\ }%
\alpha,\alpha^{\prime}=A,B,C,D,
\end{equation}
is
\begin{equation}
\label{rhode}\rho_{\alpha\alpha^{\prime}}^{DE}\left(  t\right)  =\frac
{\Gamma_{\alpha\alpha^{\prime}}}{2}\frac{e^{-\left(  \gamma_{\alpha
\alpha^{\prime}}+i\omega_{\alpha\alpha^{\prime}}\right)  t}-e^{-2\gamma_{2}t}%
}{2\gamma_{2}-\left(  \gamma_{\alpha\alpha^{\prime}}+i\omega_{\alpha
\alpha^{\prime}}\right)  }\text{ \ \ \ \ for\ }\alpha,\alpha^{\prime}=A,B,C,D,
\end{equation}
and for the SE initial condition%
\begin{equation}
\label{rhose}\rho_{EE}\left(  0\right)  =\rho_{GG}\left(  0\right)  =0;\text{
\ \ }\rho_{\alpha\alpha^{\prime}}\left(  0\right)  =1/4\text{ for\ }%
\alpha,\alpha^{\prime}=A,B,C,D,
\end{equation}
is
\begin{equation}
\rho_{\alpha\alpha^{\prime}}^{SE}\left(  t\right)  =\frac{1}{4}e^{-\gamma
_{\alpha\alpha^{\prime}}t}e^{-i\omega_{\alpha\alpha^{\prime}}t}.
\end{equation}

The corresponding values of
\begin{equation}
S_{\alpha\alpha^{\prime}}=\frac{\gamma_{2}}{c}\lim_{t\rightarrow\infty}%
\int_{R_{0}}^{ct}\rho_{\alpha\alpha^{\prime}}\left(  t-R/c\right)  dR,
\end{equation}
needed in the calculation are%
\begin{equation}
\label{esde}S_{\alpha\alpha^{\prime}}^{DE}=\frac{\Gamma_{\alpha\alpha^{\prime
}}}{4\left(  \gamma_{\alpha\alpha^{\prime}}+i\omega_{\alpha\alpha^{\prime}%
}\right)  }%
\end{equation}
and%
\begin{equation}
\label{esse}S_{\alpha\alpha^{\prime}}^{SE}=\frac{\gamma_{2}}{4\left(
\gamma_{\alpha\alpha^{\prime}}+i\omega_{\alpha\alpha^{\prime}}\right)  }.
\end{equation}

In evaluating the expectation values of operators in source-field approach, we
need the following expectation values:%
\begin{subequations}
\begin{align}
k_{11}\left(  t\right)   &  =\left\langle \sigma^{(1)}\left(  1,1,t\right)
\right\rangle +\left\langle \sigma^{(2)}\left(  1,1,t\right)  \right\rangle
=\frac{1}{2}\left[
\begin{array}
[c]{c}%
4\rho_{EE}\left(  t\right)  +\rho_{AA}\left(  t\right)  +\rho_{BB}\left(
t\right)  +\rho_{CC}\left(  t\right)  +\rho_{DD}\left(  t\right) \\
+\rho_{AB}\left(  t\right)  +\rho_{BA}\left(  t\right)  +\rho_{CD}\left(
t\right)  +\rho_{DC}\left(  t\right)
\end{array}
\right]  ,\\
k_{-1-1}\left(  t\right)   &  =\left\langle \sigma^{(1)}\left(
-1,-1,t\right)  \right\rangle +\left\langle \sigma^{(2)}\left(
-1,-1,t\right)  \right\rangle =\frac{1}{2}\left[
\begin{array}
[c]{c}%
\rho_{AA}\left(  t\right)  +\rho_{BB}\left(  t\right)  +\rho_{CC}\left(
t\right)  +\rho_{DD}\left(  t\right) \\
-\rho_{AB}\left(  t\right)  -\rho_{BA}\left(  t\right)  -\rho_{CD}\left(
t\right)  -\rho_{DC}\left(  t\right)
\end{array}
\right]  ,\\
k_{1-1}\left(  t\right)   &  =\left\langle \sigma^{(1)}\left(  1,-1,t\right)
\right\rangle +\left\langle \sigma^{(2)}\left(  1,-1,t\right)  \right\rangle
=\frac{1}{2}\left[
\begin{array}
[c]{c}%
\rho_{AA}\left(  t\right)  -\rho_{BB}\left(  t\right)  -\rho_{CC}\left(
t\right)  +\rho_{DD}\left(  t\right) \\
+\rho_{AB}\left(  t\right)  -\rho_{BA}\left(  t\right)  -\rho_{CD}\left(
t\right)  +\rho_{DC}\left(  t\right)
\end{array}
\right]  ,\\
k_{-11}\left(  t\right)   &  =\left\langle \sigma^{(1)}\left(  -1,1,t\right)
\right\rangle +\left\langle \sigma^{(2)}\left(  -1,1,t\right)  \right\rangle
=\frac{1}{2}\left[
\begin{array}
[c]{c}%
\rho_{AA}\left(  t\right)  -\rho_{BB}\left(  t\right)  -\rho_{CC}\left(
t\right)  +\rho_{DD}\left(  t\right) \\
-\rho_{AB}\left(  t\right)  +\rho_{BA}\left(  t\right)  +\rho_{CD}\left(
t\right)  -\rho_{DC}\left(  t\right)
\end{array}
\right]  ,\\
f_{11}\left(  t\right)   &  =\left\langle \sigma_{+}^{(2)}\left(  1,t\right)
\sigma_{-}^{(1)}\left(  1,t\right)  \right\rangle =\frac{1}{4}\left[
\begin{array}
[c]{c}%
\rho_{AA}\left(  t\right)  +\rho_{BB}\left(  t\right)  -\rho_{CC}\left(
t\right)  -\rho_{DD}\left(  t\right) \\
+\rho_{AB}\left(  t\right)  +\rho_{BA}\left(  t\right)  -\rho_{AC}\left(
t\right)  +\rho_{CA}\left(  t\right) \\
-\rho_{AD}\left(  t\right)  +\rho_{DA}\left(  t\right)  -\rho_{BC}\left(
t\right)  +\rho_{CB}\left(  t\right) \\
-\rho_{BD}\left(  t\right)  +\rho_{DB}\left(  t\right)  -\rho_{CD}\left(
t\right)  -\rho_{DC}\left(  t\right)
\end{array}
\right]  ,\\
f_{-1-1}(t)  &  =\left\langle \sigma_{+}^{(2)}\left(  -1,t\right)  \sigma
_{-}^{(1)}\left(  -1,t\right)  \right\rangle =\frac{1}{4}\left[
\begin{array}
[c]{c}%
\rho_{AA}\left(  t\right)  +\rho_{BB}\left(  t\right)  -\rho_{CC}\left(
t\right)  -\rho_{DD}\left(  t\right) \\
-\rho_{AB}\left(  t\right)  -\rho_{BA}\left(  t\right)  +\rho_{AC}\left(
t\right)  -\rho_{CA}\left(  t\right) \\
-\rho_{AD}\left(  t\right)  +\rho_{DA}\left(  t\right)  -\rho_{BC}\left(
t\right)  +\rho_{CB}\left(  t\right) \\
+\rho_{BD}\left(  t\right)  -\rho_{DB}\left(  t\right)  +\rho_{CD}\left(
t\right)  +\rho_{DC}\left(  t\right)
\end{array}
\right]  ,\\
f_{1-1}(t)  &  =\left\langle \sigma_{+}^{(2)}\left(  1,t\right)  \sigma
_{-}^{(1)}\left(  -1,t\right)  \right\rangle =\frac{1}{4}\left[
\begin{array}
[c]{c}%
\rho_{AA}\left(  t\right)  -\rho_{BB}\left(  t\right)  +\rho_{CC}\left(
t\right)  -\rho_{DD}\left(  t\right) \\
+\rho_{AB}\left(  t\right)  -\rho_{BA}\left(  t\right)  -\rho_{AC}\left(
t\right)  -\rho_{CA}\left(  t\right) \\
-\rho_{AD}\left(  t\right)  +\rho_{DA}\left(  t\right)  +\rho_{BC}\left(
t\right)  -\rho_{CB}\left(  t\right) \\
+\rho_{BD}\left(  t\right)  +\rho_{DB}\left(  t\right)  +\rho_{CD}\left(
t\right)  -\rho_{DC}\left(  t\right)
\end{array}
\right]  ,\\
f_{-11}\left(  t\right)   &  =\left\langle \sigma_{+}^{(2)}\left(
-1,t\right)  \sigma_{-}^{(1)}\left(  1,t\right)  \right\rangle =\frac{1}%
{4}\left[
\begin{array}
[c]{c}%
\rho_{AA}\left(  t\right)  -\rho_{BB}\left(  t\right)  +\rho_{CC}\left(
t\right)  -\rho_{DD}\left(  t\right) \\
-\rho_{AB}\left(  t\right)  +\rho_{BA}\left(  t\right)  +\rho_{AC}\left(
t\right)  +\rho_{CA}\left(  t\right) \\
-\rho_{AD}\left(  t\right)  +\rho_{DA}\left(  t\right)  +\rho_{BC}\left(
t\right)  -\rho_{CB}\left(  t\right) \\
-\rho_{BD}\left(  t\right)  -\rho_{DB}\left(  t\right)  -\rho_{CD}\left(
t\right)  +\rho_{DC}\left(  t\right)
\end{array}
\right]  ,\\
h_{11}\left(  t\right)   &  =\left\langle \sigma_{+}^{(1)}\left(  1,t\right)
\sigma_{-}^{(2)}\left(  1,t\right)  \right\rangle =\frac{1}{4}\left[
\begin{array}
[c]{c}%
\rho_{AA}\left(  t\right)  +\rho_{BB}\left(  t\right)  -\rho_{CC}\left(
t\right)  -\rho_{DD}\left(  t\right) \\
+\rho_{AB}\left(  t\right)  +\rho_{BA}\left(  t\right)  +\rho_{AC}\left(
t\right)  -\rho_{CA}\left(  t\right) \\
+\rho_{AD}\left(  t\right)  -\rho_{DA}\left(  t\right)  +\rho_{BC}\left(
t\right)  -\rho_{CB}\left(  t\right) \\
+\rho_{BD}\left(  t\right)  -\rho_{DB}\left(  t\right)  -\rho_{CD}\left(
t\right)  -\rho_{DC}\left(  t\right)
\end{array}
\right]  ,\\
h_{-1-1}\left(  t\right)   &  =\left\langle \sigma_{+}^{(1)}\left(
-1,t\right)  \sigma_{-}^{(2)}\left(  -1,t\right)  \right\rangle =\frac{1}%
{4}\left[
\begin{array}
[c]{c}%
\rho_{AA}\left(  t\right)  +\rho_{BB}\left(  t\right)  -\rho_{CC}\left(
t\right)  -\rho_{DD}\left(  t\right) \\
-\rho_{AB}\left(  t\right)  -\rho_{BA}\left(  t\right)  -\rho_{AC}\left(
t\right)  +\rho_{CA}\left(  t\right) \\
+\rho_{AD}\left(  t\right)  -\rho_{DA}\left(  t\right)  +\rho_{BC}\left(
t\right)  -\rho_{CB}\left(  t\right) \\
-\rho_{BD}\left(  t\right)  +\rho_{DB}\left(  t\right)  +\rho_{CD}\left(
t\right)  +\rho_{DC}\left(  t\right)
\end{array}
\right]  ,\\
h_{1-1}\left(  t\right)   &  =\left\langle \sigma_{+}^{(1)}\left(  1,t\right)
\sigma_{-}^{(2)}\left(  -1,t\right)  \right\rangle =\frac{1}{4}\left[
\begin{array}
[c]{c}%
\rho_{AA}\left(  t\right)  -\rho_{BB}\left(  t\right)  +\rho_{CC}\left(
t\right)  -\rho_{DD}\left(  t\right) \\
+\rho_{AB}\left(  t\right)  -\rho_{BA}\left(  t\right)  +\rho_{AC}\left(
t\right)  +\rho_{CA}\left(  t\right) \\
\rho_{AD}\left(  t\right)  -\rho_{DA}\left(  t\right)  -\rho_{BC}\left(
t\right)  +\rho_{CB}\left(  t\right) \\
-\rho_{BD}\left(  t\right)  -\rho_{DB}\left(  t\right)  +\rho_{CD}\left(
t\right)  -\rho_{DC}\left(  t\right)
\end{array}
\right]  ,\\
h_{-11}\left(  t\right)   &  =\left\langle \sigma_{+}^{(1)}\left(
-1,t\right)  \sigma_{-}^{(2)}\left(  1,t\right)  \right\rangle =\frac{1}%
{4}\left[
\begin{array}
[c]{c}%
\rho_{AA}\left(  t\right)  -\rho_{BB}\left(  t\right)  +\rho_{CC}\left(
t\right)  -\rho_{DD}\left(  t\right) \\
-\rho_{AB}\left(  t\right)  +\rho_{BA}\left(  t\right)  -\rho_{AC}\left(
t\right)  -\rho_{CA}\left(  t\right) \\
\rho_{AD}\left(  t\right)  -\rho_{DA}\left(  t\right)  -\rho_{BC}\left(
t\right)  +\rho_{CB}\left(  t\right) \\
+\rho_{BD}\left(  t\right)  +\rho_{DB}\left(  t\right)  -\rho_{CD}\left(
t\right)  +\rho_{DC}\left(  t\right)
\end{array}
\right]  .
\end{align}

The matrix elements $M_{\alpha}^{\left(  \lambda\right)  }\left(
\mathbf{k}\right)  $ needed in Eq. (\ref{examp}) are given by
\end{subequations}
\begin{subequations}
\label{matel}%
\begin{align}
M_{A}^{\left(  \theta\right)  }\left(  \mathbf{k}\right)   &  =i\frac
{1+e^{-ia}}{\sqrt{6}}\cos\theta_{k}\sin\phi_{k},\\
M_{B}^{\left(  \theta\right)  }\left(  \mathbf{k}\right)   &  =\frac
{1+e^{-ia}}{\sqrt{6}}\cos\theta_{k}\cos\phi_{k},\\
M_{C}^{\left(  \theta\right)  }\left(  \mathbf{k}\right)   &  =\frac
{1-e^{-ia}}{\sqrt{6}}\cos\theta_{k}\cos\phi_{k},\\
M_{D}^{\left(  \theta\right)  }\left(  \mathbf{k}\right)   &  =i\frac
{1-e^{-ia}}{\sqrt{6}}\cos\theta_{k}\sin\phi_{k},\\
M_{A}^{\left(  \phi\right)  }\left(  \mathbf{k}\right)   &  =i\frac{1+e^{-ia}%
}{\sqrt{6}}\cos\phi_{k},\\
M_{B}^{\left(  \phi\right)  }\left(  \mathbf{k}\right)   &  =-\frac{1+e^{-ia}%
}{\sqrt{6}}\sin\phi_{k},\\
M_{C}^{\left(  \phi\right)  }\left(  \mathbf{k}\right)   &  =-\frac{1-e^{-ia}%
}{\sqrt{6}}\sin\phi_{k},\\
M_{D}^{\left(  \phi\right)  }\left(  \mathbf{k}\right)   &  =i\frac{1-e^{-ia}%
}{\sqrt{6}}\cos\phi_{k},
\end{align}
where
\end{subequations}
\begin{equation}
a=kR_{21}\sin\theta_{k}\cos\phi_{k}.
\end{equation}

In carrying out the integrations in Eq. (\ref{angtot}) in WWA, we encounter
integrals of the type
\[
\int_{-\infty}^{\infty}d\omega_{k}\frac{j_{n}\left(  \omega_{k}R_{21}%
/c\right)  }{\left[  \gamma_{\alpha}/2-i\left(  \omega_{k}-\omega_{\alpha
}\right)  \right]  \left[  \gamma_{\alpha^{\prime}}/2+i\left(  \omega
_{k}-\omega_{\alpha^{\prime}}\right)  \right]  },
\]
where $j_{n}$ ($n=0,1,2,3$) is a spherical Bessel function. Consistent with
the WWA and inequalities (\ref{shifts}) and (\ref{retard}), we replace this
integral by%
\[
2\pi\frac{j_{n}\left(  k_{0}R_{21}\right)  }{\gamma_{\alpha\alpha^{\prime}%
}+i\omega_{\alpha\alpha^{\prime}}}.
\]


\begin{thebibliography}{99}
\bibitem {math}Mathematica notebooks used in the calculations are available at xxxxxx.
\bibitem {torque}Alternatively, the torque can be calculated as $\mathbf{R}%
_{2}\times\mathbf{F}_{21}(\mathbf{R}_{2},t)=2\mathbf{R}_{2}\times
\operatorname{Re}\left\langle \boldsymbol{\nabla}\left[  \boldsymbol{\mu}%
_{+}\cdot\mathbf{E}_{s}^{+}(\mathbf{R},t)\right]  \right\rangle _{\mathbf{R}%
=\mathbf{R}_{2}}$, where $\boldsymbol{\mu}_{+}=e^{i\omega_{0}t}\sum_{m=-1}%
^{1}\boldsymbol{\mu}_{-}^{\ast}\left(  m\right)  \sigma_{+}^{(2)}(m,t)$, or by
fully quantizing the center of mass motion and calculating $d\left\langle
\mathbf{L}\right\rangle /dt$ associated with the relative motion of the two atoms.
\end{thebibliography}
\end{document}